\documentclass[aps,pra,twocolumn,groupedaddress,showpacs,10pt]{revtex4-1}
\usepackage{graphicx}
\usepackage{amsfonts}
\usepackage{color}

\begin{document}



\title{Four-Color Stimulated Optical Forces for Atomic and Molecular Slowing}

\author{S. E. Galica}
\author{L. Aldridge}
\author{E. E. Eyler}
\affiliation{Physics Department, University of Connecticut, Storrs, CT 06269}

\date{\today}

\begin{abstract}

Stimulated optical forces offer a simple and efficient method for providing optical forces far in excess of the saturated radiative force. The bichromatic force, using a counterpropagating pair of two-color beams, has so far been the most effective of these stimulated forces for deflecting and slowing atomic beams.  We have numerically studied the evolution of a two-level system under several different bichromatic and polychromatic light fields, while retaining the overall geometry of the bichromatic force.  New insights are gained by studying the time-dependent trajectory of the Bloch vector, including a better understanding of the remarkable robustness of bi- and polychromatic forces with imbalanced beam intensities.  We show that a four-color polychromatic force exhibits great promise.  By adding new frequency components at the third harmonic of the original bichromatic detuning, the force is increased by nearly 50\% and its velocity range is extended by a factor of three, while the required laser power is increased by only 33\%.  The excited-state fraction, crucial to possible application to molecules, is reduced from 41\% to 24\%.  We also discuss some important differences between polychromatic forces and pulse trains from a high-repetition-rate laser.
\end{abstract}

\pacs{37.10.Vz, 37.10.De}

\maketitle

\section{Introduction}

Since their inception, optical forces have become an invaluable tool in atomic and molecular physics, playing a dominant role in the preparation and study of ultracold neutral species. These forces can be divided into two main categories, spontaneous and stimulated. As the names imply, spontaneous optical forces rely on spontaneous decay of a system to accomplish momentum transfer, while stimulated forces make use of the surrounding light field to drive both absorption and emission. Stimulated forces have the advantage of allowing much greater forces at the cost of more complex optical configurations.

Previous experiments have demonstrated the effectiveness of stimulated forces, which rely on coherent momentum transfer between a system (e.g. an atom) and the light field. Grimm and coworkers made use of rectified optical dipole forces to deflect an atomic Cs beam by several m/s ~\cite{Grimm90}, later improving on this result by developing the much stronger optical bichromatic force (BCF) and using it to decelerate the atoms nearly to rest \cite{Soding97}. The basic principles of the BCF, which utilizes two-color beams with symmetric detunings $\pm \delta$ from resonance, are reviewed in Sec. \ref{sec:BCF} below.  Further work by Metcalf expanded on this idea by using bichromatic forces to focus and decelerate an atomic beam of metastable helium (He*) \cite{Williams99, Williams00, Partlow04}. More recently, our group has designed and evaluated two schemes for longitudinal slowing of a beam of metastable helium by several hundred m/s ~\cite{Chieda12, ChiedaThesis}. In one approach, an amplified laser was used at a large detuning from resonance to extend the velocity range over which the force is effective, but it became clear that a complicated high-power multi-stage design would be needed. In the other, deceleration by nearly 400 m/s was achieved using two ordinary diode lasers in a chirped bichromatic slowing configuration that compensated the changing Doppler shifts.  As described in Ref. \cite{Chieda12}, an upgraded version using slightly larger laser detuning and higher power is predicted to slow He* to velocities suitable for MOT loading.  Enhanced MOT loading using BCF slowing has already been demonstrated for atomic Rb, for which Doppler shifts are of much less concern \cite{Liebisch12}.

The most important limits constraining the bichromatic force are the magnitude of the force, the velocity range, and the excited-state fraction.  Though it might seem counter-intuitive, the magnitude of the force is perhaps the least important of these, simply because it is so large.  Under typical BCF conditions the stopping distance for a typical atomic beam would be on the order of 1 cm if a constant force could be maintained.  Instead, the attainable velocity reduction is usually limited either by the velocity range of the force or by losses unrelated to the BCF cycle itself, a particular concern if the BCF is to be applied to molecular systems~\cite{Chieda11}.  For molecules, there is a strong motivation to reduce the excited-state fraction in order to reduce radiative losses into ``dark" states with the wrong vibrational or rotational quantum number.  These losses effectively terminate the BCF cycle and thereby limit the time interval available for deceleration.

In this paper we describe detailed numerical calculations for a two-level atom or molecule subjected to bichromatic or polychromatic laser beams in the BCF configuration --- a pair of counterpropagating multicolor beams with an adjustable phase shift. The forces and ensemble behavior due to stimulated optical forces are calculated by direct numerical solution of the optical Bloch equations (OBEs).  We pay particular attention to the time evolution of atomic excitation, which has not previously been investigated in a way that allows systematic examination. We also carefully investigate the robustness of bichromatic and polychromatic forces against imbalances in intensity between counterpropagating beams. Insensitivity to beam imbalance is vital for successful experimental realizations because an intensity balance of less than 5-10\% is very difficult to attain over an extended area.

We begin with a discussion of the conventional two-color bichromatic force, with an emphasis on factors that define its limits as a practical slowing mechanism. An unusual ``Bloch cylinder" plot is introduced to graphically visualize some of the more obscure aspects of bichromatic forces. We then discuss a proposed extension to multicolor or polychromatic forces (PCFs), and more specifically to a four-color stimulated force. Heuristic arguments based on pulse areas suggest that additional rf sidebands at $\pm 3\delta$ should reduce the excited-state fraction while maintaining a large optical force, and this prediction is verified by the actual modeled behavior.  We show that for an increase of just 33\% in the total laser power, large improvements can be obtained not only in the excited-state fraction, but also in the magnitude and velocity range of the force. The projected improvements are sufficiently large to make the four-color scheme attractive both for molecules and for atomic beam deceleration.  We also briefly discuss considerations that arise in the many-color limit of a continuous train of short pulses.

\section{Bichromatic Force \label{sec:BCF}}

\begin{figure}
\includegraphics[width=\linewidth]{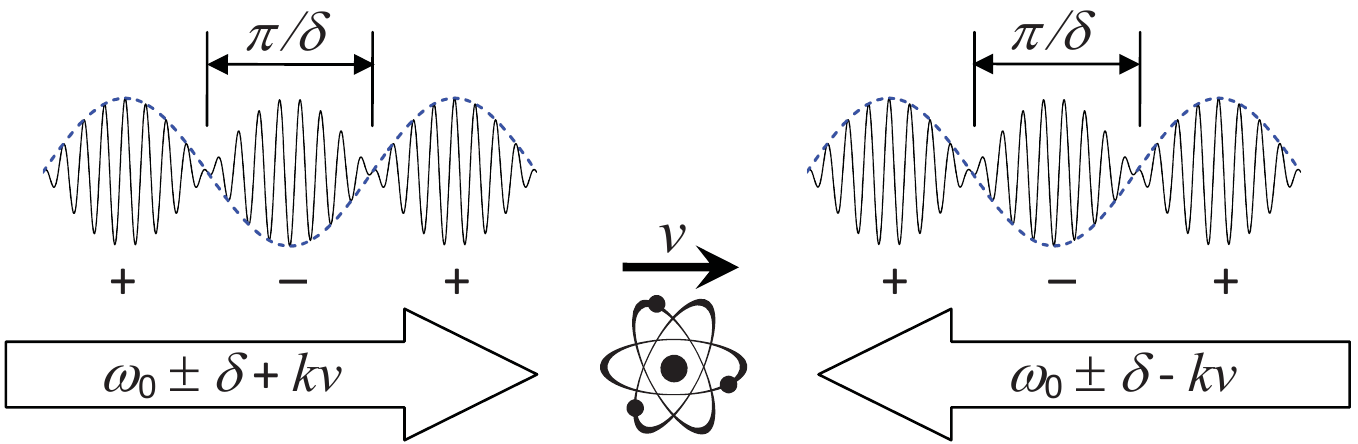}
\caption{Simple $\pi$-pulse model of the bichromatic force on a two-level atomic or molecular system, based on Fig. 1 in Ref. \cite{Chieda12}. From each direction, a pair of beams symmetrically detuned from resonance by rf frequencies $\pm \delta$ interfere to form beat notes, each with an area of approximately $\pi$. Because the beat note envelope (blue dashed line) alternates in sign as shown, the Rabi frequency exhibits similar sign reversals.  For high velocities, a Doppler offset of $\pm kv$ may be necessary to bring the system within the velocity range of the force.}
\label{fig:BCF_concept}
\end{figure}

As the name suggests, the bichromatic force is a particular realization of a two-color stimulated optical force, first proposed in the late 1980s by Voitsekhovich, \emph{et al.}~\cite{Voitsekhovich89}. It uses counterpropagating pairs of two-color laser beams to stimulate excitation and emission in an ensemble of atoms or molecules. As shown in Fig. \ref{fig:BCF_concept}, each beam has components detuned by rf frequencies $\pm\delta$ from resonance. These near-resonant beams interfere to form beat notes with a period $\pi/\delta$ as illustrated.  Because the full period of the beat note envelope is actually $2\pi/\delta$, there are phase reversals between adjacent pulses, which will play an important role in our discussion of the robustness of bichromatic and polychromatic forces.

The intensity of the beat notes is adjusted such that each ``pulse" of duration $\pi/\delta$ has an area of approximately $\pi$ in units normalized to the Rabi frequency \cite{Milonni10}.  In a simplified model of the BCF usually called the $\pi$-pulse model, the pulses from the right and the left are regarded as if they were non-overlapping short pulses, so that an atom at the center experiences alternating $\pi$-pulses that cause a repeating cycle of excitation from the right followed by stimulated emission from the left.  Each excitation or stimulated decay produces a momentum change $\hbar k$, at a rate $\delta/\pi$ set by the beat frequency.  This will continue until radiative decay randomly resets the atom into the ground state.  By adjusting the phase between the counterpropagating beat-note trains from the right and the left, one can control the probability that radiative decay will restart the system in the correct right-left sequence, thereby controlling the average direction of the applied force.  The maximum decelerating force on an ensemble of atoms occurs when the phase shift between beats corresponds to half of a beat note or $\chi = \pi/4$, where $\chi$ is the phase of the electric field envelope.  This optimal phasing yields a time-averaged bichromatic force on each atom of
\begin{equation}\label{eq:F_bich}
F_{\text{bich}} = \frac{\hbar k \delta}{\pi}.
\end{equation}
These results for the force and the optimal phasing remain valid in more accurate treatments using doubly-dressed atoms or numerical methods, although the optimal pulse area for each beat note increases to $1.559\pi$ when the overlapping beats are fully taken into account \cite{Cashen03,Yatsenko04,ChiedaThesis}. The optimization of these parameters is explored in Refs. \cite{Chieda12, ChiedaThesis,Yatsenko04}, together with discussions of the sensitivity the BCF to deviations from the optimal values. However, the effects of imbalanced intensities between counterpropagating beam pairs have not been well-understood, and we address this topic below.

\subsection{Numerical solution of the Optical Bloch Equations \label{subsec:OBEsim}}

To systematically study the evolution of an ensemble over time, both for the BCF and its polychromatic generalizations, the OBEs for a two-level atom are solved in the rotating-wave approximation for a multi-color light field. This direct numerical approach, originally developed by S\"{o}ding and coworkers \cite{Soding97}, has proven quite reliable for modeling the bichromatic force even in the presence of dynamically changing conditions \cite{Chieda12}. We solve the OBEs in the form \cite{ChiedaThesis}
\begin{eqnarray}
\label{eq:OBEs}
&&\frac{du}{dt}  =  -\gamma u - \delta_{\text{asym}} v - \text{Im}\left[\Omega(t)\right] w \nonumber \\
&&\frac{dv}{dt}  =  \delta_{\text{asym}} u - \gamma v + \text{Re}\left[\Omega(t)\right] w \\
&&\frac{dw}{dt}  =  \text{Im}\left[\Omega(t)\right] u - \text{Re}\left[\Omega(t)\right] v -2\gamma (w+1), \nonumber
\end{eqnarray}
where $\gamma$ is the excited-state lifetime and $\Omega(t)$ is the time-varying Rabi frequency.  In the present context the parameter $\delta_{\text{asym}}$ is useful mainly to allow modeling of asymmetric detunings --- it is the shift of the optical carrier frequency relative to the atomic resonance frequency.  The Bloch vectors $u$, $v$, and $w$ can be expressed in terms of the density matrix $\rho_{ij}$,
\begin{eqnarray}
&&u  =  \rho_{12}+\rho_{21}, \nonumber \\
&&v  =  i(\rho_{12}-\rho_{21}), \\
&&w  =  \rho_{22}-\rho_{11}. \nonumber
\end{eqnarray}
Thus $w=-1$ corresponds to a pure ensemble of ground-state atoms and $w=+1$ to the excited state.

It remains to determine the Rabi frequency $\Omega$ in Eqs. (\ref{eq:OBEs}) for the multiple laser fields used in the BCF and its PCF generalization.  For a resonant BCF configuration with symmetrically detuned bichromatic fields incident from each direction, the total electric field is a sum of two beat-note trains,
\begin{eqnarray}
\label{eq:EFieldsBal}
E(z,t) & = & 2E_0 \cos\left[\omega(t-z/c) \right]
\cos\left[\delta(t-z/c)+ \chi/2 \right] \nonumber\\
  & + & 2E_0\cos\left[\omega(t+z/c) \right]
 \cos\left[\delta(t+z/c) - \chi/2 \right] \nonumber,\\
\end{eqnarray}
where each field component has amplitude $E_0$, the detunings from resonance are $\pm \delta$, and $\chi$ is the left-vs-right phase shift discussed previously in association with Fig. \ref{fig:BCF_concept}.  Assuming that the beat note length is large compared to the size of the laser interaction region, $c/\delta \gg \Delta z$, the dependence on $\omega$ can be factored out to yield

\begin{eqnarray}
\label{eq:EFieldsApprox}
E(z,t)& \cong & 2E_0 e^{i\omega t} \left[\cos(k z)\cos(\delta t)\cos(\chi/2)\right. \nonumber\\
& + & \left. i\sin(k z)\sin(\delta t)\sin(\chi/2)\right] + \text{c.c.},
\end{eqnarray}
where $k=\omega/c$, as is customary.

In the rotating-wave approximation (RWA), the corresponding Rabi frequency is readily found from Eq. \ref{eq:EFieldsApprox} and the electronic dipole matrix element $\mu$,
\begin{eqnarray}
\label{eq:RabiFreq}
\Omega^{\text{BCF}}(t)& = & \frac{4\mu_e E_0}{\hbar} [ \cos(kz) \cos(\delta t) \cos(\chi/2) \nonumber\\
& + & i \sin(kz) \sin(\delta t ) \sin(\chi/2) ] .
\end{eqnarray}
This expression can easily be extended to polychromatic fields with detunings $\pm\delta$, $\pm2\delta$,..., and real amplitudes $E_n$.  The Rabi frequency then becomes
\begin{eqnarray}
\label{eq:RabiPCF}
\Omega(t)& = & \frac{4\mu_e}{\hbar} \sum_{n=1}^{n_{\text{max}}} E_n \left[ \cos(kz)
 \cos(n\delta t + \theta_n ) \cos(n\chi/2) \right. \nonumber\\
& + & \left. i \sin(kz) \sin(n\delta t + \theta_n ) \sin(n\chi/2) \right],
\end{eqnarray}
 where the additional phases $\theta_n$ are harmonic phases that define the Fourier superposition of the frequency components $n\delta$. For convenience in the discussions that follow, we collect the factors defining the Rabi frequency for each harmonic $n$ as $\Omega_n\equiv\mu_eE_n/\hbar$.  We note that for the ordinary BCF with $n$=1, the $\pi$-pulse condition can be expressed as $\Omega_1 = (\pi/4)\delta$, and the optimal value is slightly larger at $\Omega_1 = \sqrt{3/2}\, \delta$ \cite{Williams00,Chieda12}.

Our computer models use slightly extended versions of Eqs. (\ref{eq:RabiFreq}) and (\ref{eq:RabiPCF}) that can also accommodate Doppler shifts $\Delta \omega = \pm kv$ for moving atoms.  In addition, we can optionally incorporate left-vs-right intensity imbalances by assigning separate electric field amplitudes $E_L$ and $E_R$ depending on the direction of incidence. We have solved the resulting OBEs (Eqs. (\ref{eq:OBEs})) using both a standard FORTRAN ODE solver \cite{Shampine75} and the built-in differential equation solver in Mathematica Version 8.  Apart from fine-grained numerical noise, the agreement is excellent in all cases that we tested, inspiring confidence in the stability and accuracy of the numerical solutions.

\begin{figure}
\includegraphics{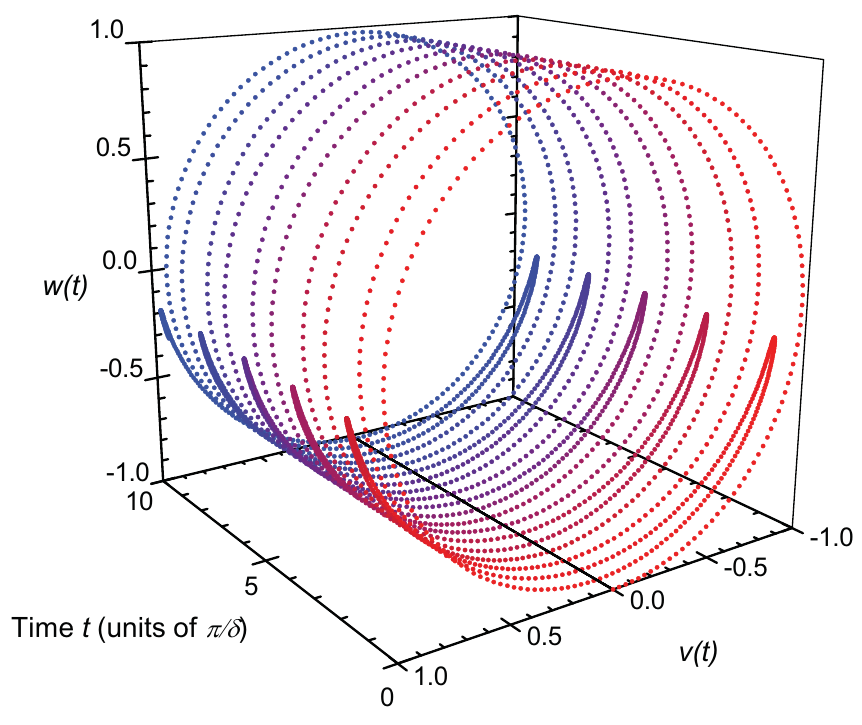}
\caption{(Color online) The evolution of the Bloch vector for a two-level system in an optimally configured bichromatic light field. The system is initialized in the ground state and allowed to evolve without radiative damping. This plot was computed for a bichromatic detuning of $\delta=125\gamma$ and deceleration parameters of $\chi=\pi/4$ and $\Omega_1=\sqrt{3/2}\:\delta$.}
\label{fig:BCF_ensemble}
\end{figure}

For any fixed location, the coordinate origin can be selected so that $z=0$ and the Rabi frequency is purely real. If the calculations are initialized in the ground state and the bichromatic detuning asymmetry $\delta_{\text{asym}}$ is zero, the first of Eqs. \ref{eq:OBEs} indicates that the $u$-component of the Bloch vector remains close to zero for short times, $\gamma t \ll 1$.  We can thus disregard the behavior of the $u$-component to study the behavior of the stimulated cycling for short times, allowing us to plot the trajectory of the Bloch vector in the $v$-$w$ plane as a function of time.  An example of the resulting ``Bloch cylinder" plot calculated under optimal BCF conditions is shown in Fig. \ref{fig:BCF_ensemble}, and reveals some aspects of the Bloch vector evolution that were previously not clearly understood.

The most striking feature is a clear deviation from the simplified model of alternating $\pi$-pulses that act in pairs to cycle the atomic excitation. Rather than simply oscillating between the poles of the Bloch sphere (as is suggested by the $\pi$-pulse model), the Bloch vector initially rotates counterclockwise through nearly $5\pi/2$ in response to an initial pair of beat notes, then reverses direction and follows a similar trajectory backwards. The additional wrapping of the Bloch vector is symmetric about the ground state of the system, but these excursions illustrate an increase in the excited-state fraction for the ensemble as compared to an idealized short-pulse sequence.  The overall four-pulse periodicity, which has apparently not previously been noted, stems from the phase reversal between successive pulses in the beat-note trains, as indicated by the (+) and (--) signs in Fig. \ref{fig:BCF_concept}.  This causes the corresponding Rabi frequencies to alternate in sign, or equivalently the pulse areas can be said to alternate between positive and negative values.  As we discuss below, the full reversal of the Bloch vector evolution after four pulses helps to explain why the BCF is highly robust in the presence of imbalanced beam intensities, both in numerical calculations and in laboratory experiments.

\begin{figure}
\includegraphics[width=0.9\linewidth]{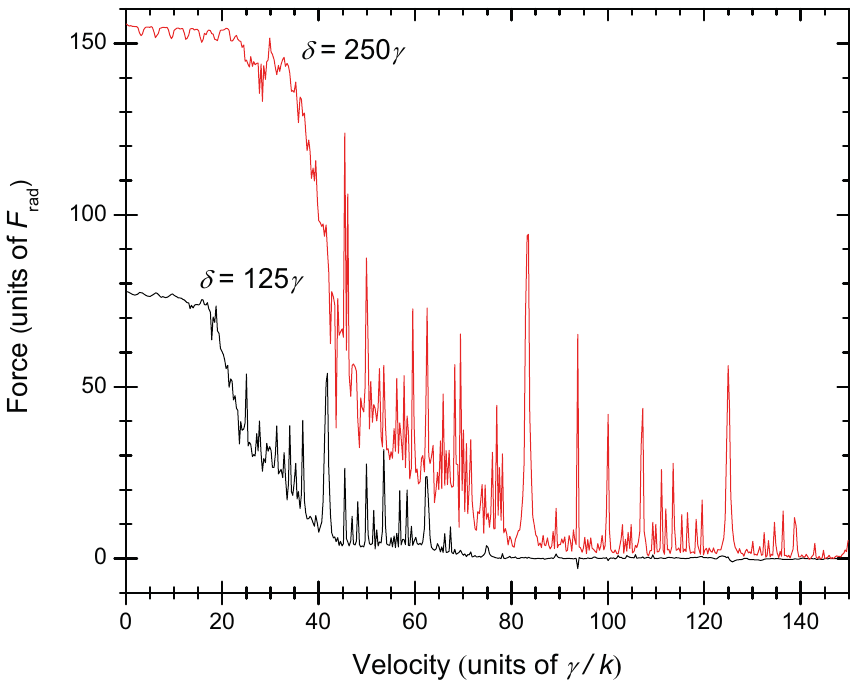}
\caption{(Color online) Numerically calculated force profiles as a function of velocity at two values of the bichromatic detuning $\delta$, using optimal values for $\chi$ and $\Omega_1$ as in Fig. \ref{fig:BCF_ensemble}. The vertical axis is scaled in units of the radiative force $F_{\text{rad}}$ and the velocity in units of $\gamma/k$. The magnitude and width of the force both scale linearly with $\delta$.  The many sharp spikes are multiphoton resonances and are typically unobservable (see text).}
\label{fig:BCF_force}
\end{figure}

In addition to modeling the ensemble behavior, solving the OBEs for atoms with fixed non-zero velocities allows us to calculate a force profile for the BCF as a function of velocity.  As in previous work \cite{Soding97,Chieda12}, the force is calculated at each velocity by use of Ehrenfest's theorem.  The force profiles shown in Fig. \ref{fig:BCF_force} exhibit the familiar features of the BCF: the force is proportional to the detuning, and so is the velocity range over which it is effective.  The horizontal axis is in scaled units of $\gamma/k$, which for the example of He* corresponds to units of 1.755 m/s, and the vertical axis is in units of the saturated radiative force from a monochromatic beam, $F_{\text{rad}}=\hbar k \gamma/2$.   Because the stimulated cycling rate depends on $\delta$ rather than $\gamma$, the magnitude of the bichromatic force is typically far in excess of the radiative force, by a factor of about $2\delta/(\pi\gamma)$.  The numerous narrow spikes in Fig. \ref{fig:BCF_force} are not numerical noise, but higher-order resonances somewhat related to ``doppleron" resonances~\cite{Bigelow90,Tollett90}. They are typically unobservable under experimental conditions due to decoherence and deviations from pure two-level behavior.

From calculations like those shown in Fig. \ref{fig:BCF_force}, the magnitude of the force is found to be in agreement with Eq. \ref{eq:F_bich}, and the full width of the velocity profile (including negative velocities) can be estimated to be~\cite{Chieda12}
\begin{equation}
\Delta v \simeq \frac{\delta}{k}.
\end{equation}
So long as the beat notes have the optimal pulse area, the force will increase in proportion to an increasing bichromatic detuning $\delta$. However, to maintain the required pulse area the Rabi frequency must also be increased proportionally to $\delta$, so the required laser intensity increases quadratically.  More specifically, the required intensity in each detuned component beam is given by
\begin{equation}
I_b=2I_s \left(\frac{\Omega_n}{\gamma}\right)^2,
\end{equation}
where $\Omega_n$ is linear in $\delta$ (e.g. $\Omega_1=\sqrt{3/2}\:\delta$ in the case of ideal BCF) thereby making very large detunings impractical.

\begin{figure}
\includegraphics{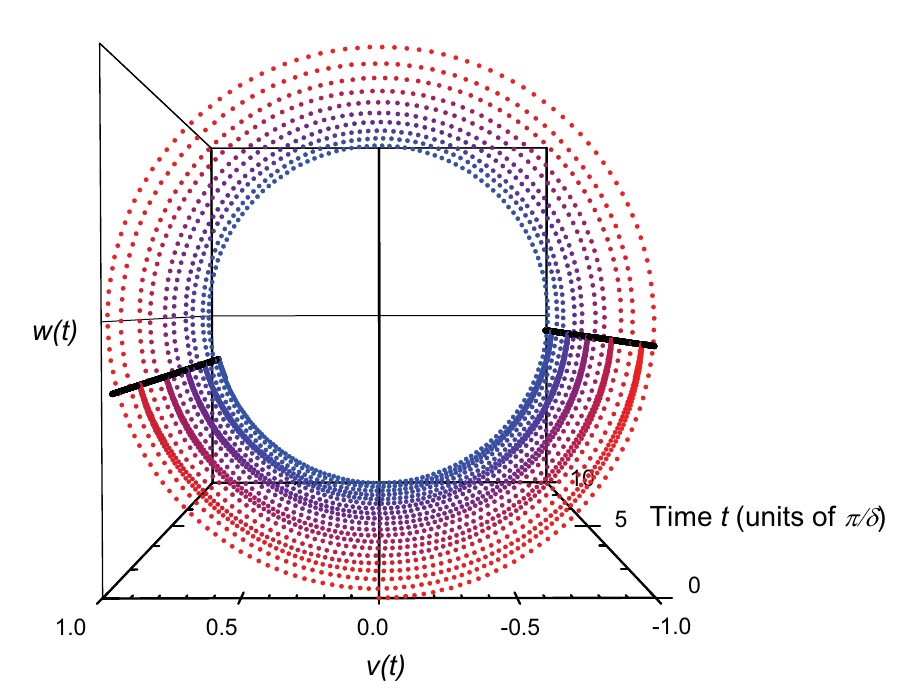}
\caption{(Color online) End-on view of the Bloch cylinder for an intensity imbalance of 25\% between bichromatic beams from the left and the right.  The thick straight lines show the turning points of the Bloch vector trajectory.  Here the axis of symmetry shifts appreciably to the right of the ground state at $w=0$, but there is still no accumulating offset of the Bloch vector phase.}
\label{fig:BCF_ensembleImb}
\end{figure}

We have previously found that as the detuning increases there is increased sensitivity to imbalanced beam intensities~\cite{Chieda12}.  However, numerical modeling showed much less sensitivity than a simple estimate based on the $\pi$-pulse model using pulse pairs, and we also noted that the simple estimate appeared to overstate the sensitivity by a factor of two for the only case in which direct comparison with experiment is possible~\cite{Soding97}.

\begin{figure}
\includegraphics[width=0.9\linewidth]{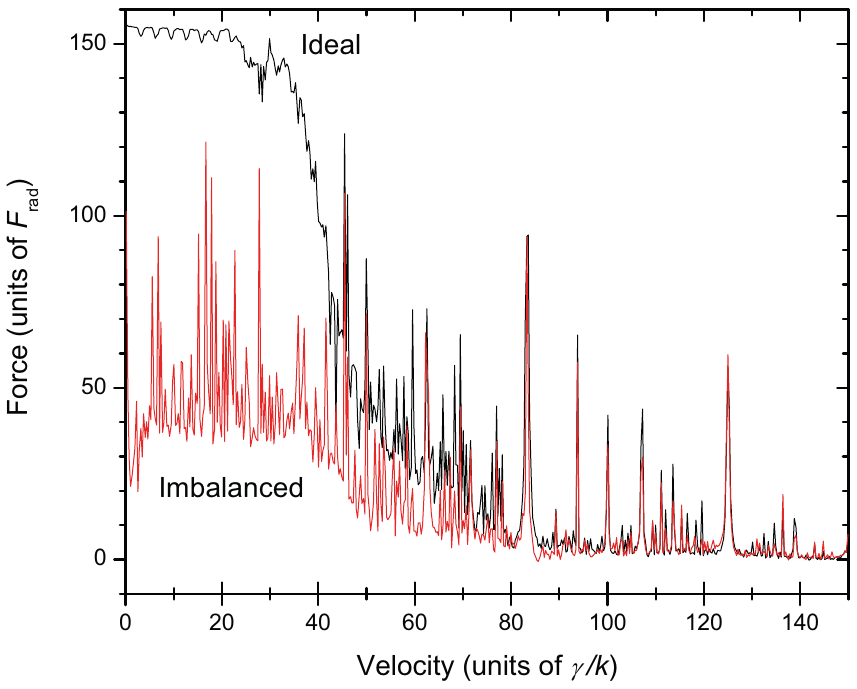}
\caption{(Color online) Top: BCF velocity profile for ideal parameter values at detuning $\delta=250\gamma$ (same as upper trace in Fig. \ref{fig:BCF_force}) Bottom: velocity profile for a left-right intensity imbalance of 25\%.}
\label{fig:BCF_ImbForceVsVelocity}
\end{figure}
\begin{figure}
\includegraphics[width=0.9\linewidth]{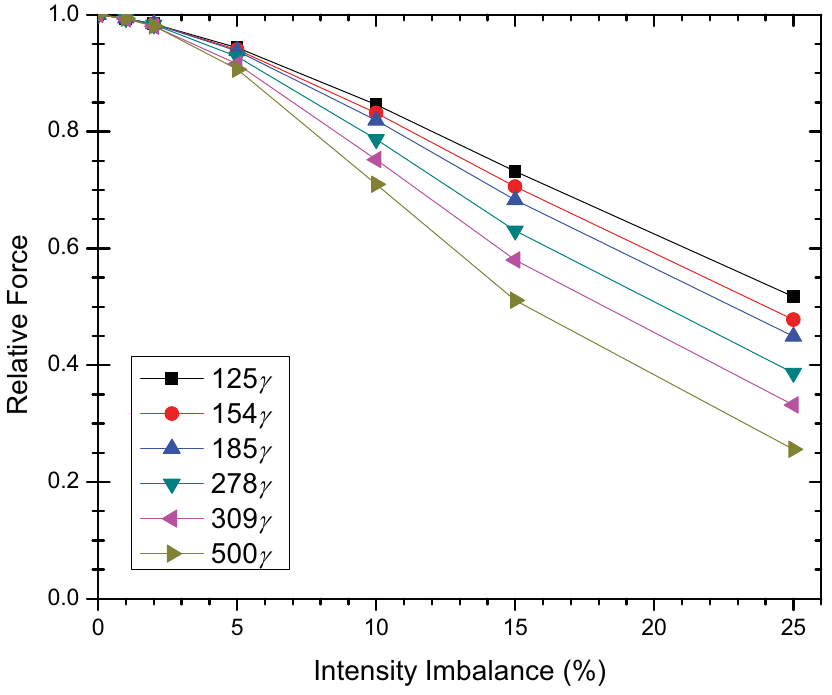}
\caption{(Color online) As imbalance between pulses increases, the force slowly degrades, at a rate that increases as the rf detuning $\delta$ is increased from 125$\gamma$ to 500$\gamma$.}
\label{fig:BCF_ForceVsImb}
\end{figure}

Much of this can be explained by examining the Bloch cylinder plots for calculations using imbalanced beams.  In Fig. \ref{fig:BCF_ensembleImb}, the Bloch vector trajectory is plotted for a left-right intensity imbalance of 25\%, where imbalance is defined as $(I_{\text{left}} - I_{\text{right}})/I_{\text{right}}$.  In the simplified $\pi$-pulse model there would be a shift in the angle of the Bloch vector after each pulse pair, resulting in a cumulative error that would reverse the sign of the force after several cycles.  Instead we see that the Bloch vector continues to execute a repetitive cycle, since a complete group of four pulses still has a total pulse area of zero.  However, the center of symmetry is slightly shifted away from the ground state, as evidenced by the asymmetric locations of the turning points in the figure.  This slightly increases the excited state fraction, but more importantly, it greatly affects the force profile.  As shown in Fig. \ref{fig:BCF_ImbForceVsVelocity}, the average force is reduced by about a factor of about three under these conditions.  A more complete picture of the force reduction force can be gained from Fig. \ref{fig:BCF_ForceVsImb}, where the relative force as a function of intensity imbalance is shown for several values of the bichromatic detuning.  For large imbalances at large detunings, examination of the force profiles reveals that they are increasingly dominated by the ``hash" of very narrow multiphoton resonances.  Because the presence of these resonances relies on a high level of coherence in a pure two-level system, the calculations in this range are expected to overestimate the force that could be realized in actual experimental conditions.

\subsection{Excited-state fraction \label{subsec:BCF_ExcFrac}}

As mentioned previously, the excited-state fraction is of particular interest because it determines the ensemble-averaged radiative decay rate.  For molecular systems a small fraction of radiative decays will unavoidably ``leak" to dark states from which further cycling cannot occur, so the time interval during which the BCF can decelerate molecules is inversely proportional to the excited-state fraction~\cite{Chieda11}.  If this fraction can be reduced while keeping the magnitude of the force constant, the velocity change attainable in a decelerator will increase correspondingly.

This fraction can easily be calculated from the time evolution of the Bloch vector.  Unlike the results discussed so far, it is important to include the effects of radiative damping by integrating Eqs. \ref{eq:OBEs} over a sufficiently large time interval, $\gamma t \gg 1$.  For BCF deceleration under optimal conditions, the calculated excited-state fraction is 41\%, independent of the detuning $\delta$.  This is somewhat smaller than the value of nearly 50\% for the ordinary radiative force on a two-level system, because the radiative force normally involves strongly saturated near-resonant excitation while the BCF utilizes large detunings.

It is also interesting to compare this result with the value predicted by careful application of the isolated $\pi$-pulse model.  Assuming an optimal left-vs.-right phase $\chi = \pi/4$, and that the atom is cycling in the correct sequence to produce deceleration as in Fig. \ref{fig:BCF_concept}, the atom spends 1/4 of its time in the excited state.  But once it radiatively decays during one of these excited-state intervals, the cycle is reversed and the atom spends 3/4 of its time in the excited state until it again decays, resuming the original cycling.  The reversed cycle is short-lived compared to the correct one because of its large excited-state fraction.  Taking a properly weighted average over these cycling conditions, the estimated excited-state fraction is 37.5\%, which is remarkably close to the exact value of 41\% calculated numerically for actual BCF conditions.

The $\pi$-pulse perspective suggests that if beat-note pulse pairs from the left and the right could be moved closer together without disrupting the cycling, a reduced excited-state fraction would result.  It was this idea that initially motivated us to consider a polychromatic variation in which additional rf sidebands at $\pm n \delta$ are introduced to produce a shorter-duration beat note pulse, allowing closer pulse proximity without excessive overlap.

\section{Polychromatic Forces \label{sec:PCF}}

As discussed in the previous section, a reduction of the excited-state fraction requires a reduction in the time between excitation and de-excitation pulses, which requires the production of narrower, better-separated pulses.  This can be done by adding additional rf sidebands to the bichromatic beam, allowing a measure of control over the pulse shape and the timing of sequential pulses.  Our hope was that this could allow not only a reduction in the left-right phase offset $\chi$ in Eq. \ref{eq:RabiPCF}, but also an opportunity to use the new adjustable parameters to optimize the characteristics of the stimulated force, in a simple version of coherent control.

\subsection{4-Color Forces \label{subsec:4color}}

In principle, a train of delta-function pulses could be produced by adding an infinite number of odd harmonics of the detuning $\pm \delta$. We thus started our investigations by adding symmetric components at $\pm 3 \delta$, although we have subsequently also briefly studied the effects of adding additional components at $\pm 2 \delta$ or $\pm 5 \delta$.  One might expect that these new third-harmonic components should be added in phase with the fundamental components at $\delta$ and with the same amplitude, since this corresponds to the lowest-order approximation to a delta function.  However, this requires systematic confirmation. It is not trivial to optimize the stimulated force with these additions, because the phase shift $\chi$ and the Rabi frequencies $\Omega_n$ must be re-optimized each time a change is made.

Numerically solving the optical Bloch equations over a coarse parameter space spanning pulse phases $\chi \in \left[0,\pi\right]$ and Rabi frequencies $\Omega_n \in \left[0,2\delta\right]$ (where $\delta$ is again the principal rf detuning) yielded an excitation fraction minima in the region of $\Omega_{1}$\,=\,$\Omega_{3}$\,=\,$\delta$ and $\chi\approx\pi/6$. The force was then calculated with finer granularity in this region.  The qualitatively best compromise between force width, force magnitude, and excited state fraction was determined to lie at $\Omega_{1}$\,=\,$\Omega_{3}$\,=\,$\delta$ and $\chi=\pi/6$ with an average excited state fraction of $24\%$. This is a 41\% reduction compared to the ordinary BCF.  We can also compare this to the predictions of the isolated $\pi$-pulse model.  Using the same argument as in Section \ref{subsec:BCF_ExcFrac} with a phase of $\pi/6$, the predicted excited-state fraction is $27.8\%$. This is again very close to the numerical calculation, but unlike the bichromatic case the numerical value is actually slightly better than the $\pi$-pulse estimate.

\begin{figure}
\includegraphics{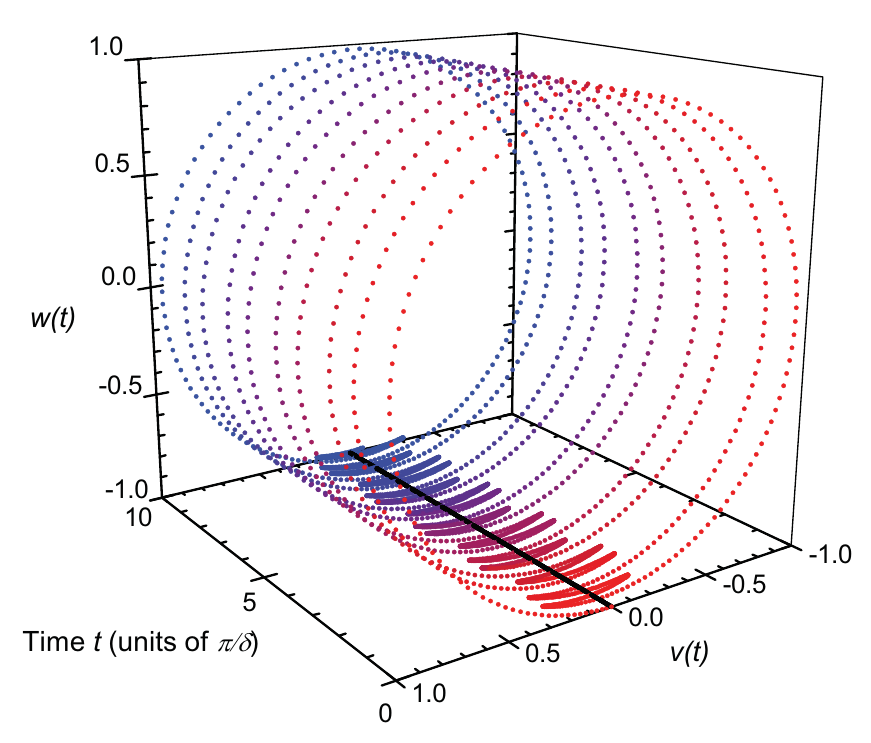}
\caption{(Color online) The evolution of the Bloch vector for a two-level system in a four-color light field. The system is initialized in the ground state and allowed to evolve under ideal deceleration conditions. Note the behavior is much closer to the $\pi$-pulse model than the bichromatic case, with the notable exception that the overall periodicity still involves four pulses, not two. The excited state character of the system is greatly reduced compared to the BCF case in Fig. \ref{fig:BCF_ensemble}. This plot was computed for a fundamental detuning of $\delta = 125 \gamma$ and deceleration parameters of $\chi = \pi/6$ and $\Omega_1 = \Omega_3 = \delta$. }
\label{fig:PCF_ensemble}
\end{figure}

The reasons for the large improvement in excited-state fraction are evident when we examine the ensemble evolution for the four-color force in a Rabi cylinder plot, as shown in Fig. \ref{fig:PCF_ensemble}.
As expected, the four-color field generates ensemble behavior more closely resembling that of a pair of separated $\pi$-pulses than the bichromatic case in Fig. \ref{fig:BCF_ensemble}. In particular, there are two key differences. First, the trajectory wobbles much less around the ground state during the low-intensity portions of the cycle. This of course translates to significantly less excited-state character in the system. Second, we note that successive points on the trajectory plots are separated by equal-sized time steps. Hence one can extract qualitative information on the velocity of the Bloch vector. Comparing with the bichromatic ensemble behavior, the four-color force moves much more quickly through the excited state (sparse dots) than its two-color counterpart, and then much more slowly as it lingers near the ground state (dense dots). These considerations also reduce the excited-state fraction relative to the BCF.

\begin{figure}
\includegraphics[width=0.9\linewidth]{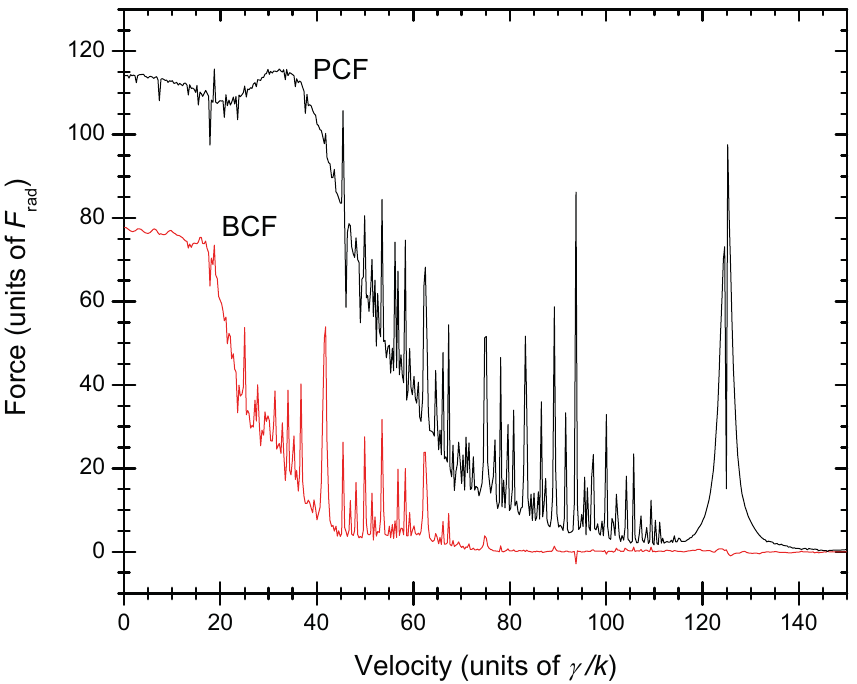}
\caption{(Color online) The forces due to bichromatic and four-color light fields at their respective optimal deceleration conditions.  The width of the four-color force is nearly equal to the width of a bichromatic force with triple the rf detuning, but at only $15\%$ of the laser power that it would require.}
\label{fig:PCF_force}
\end{figure}
The four-color force also offers a much-improved velocity range, as well as a significantly increased magnitude.  As shown in Fig. \ref{fig:PCF_force}, the maximum force magnitude in the 4-color case is increased by roughly $50\%$ over the bichromatic force. This increase is notable, but not too surprising given the 33\% increase in total laser power required to generate the four-color pulse trains. The more remarkable feature of the four-color force is its extremely wide velocity range, increased by nearly a factor of three relative to the BCF. In figure 8, the $125\gamma$-detuned four-color force is able to achieve the static range of the $375\gamma$-detuned bichromatic force at roughly 15\% of the total laser power that the BCF would require.

\begin{figure}
\includegraphics[width=0.9\linewidth]{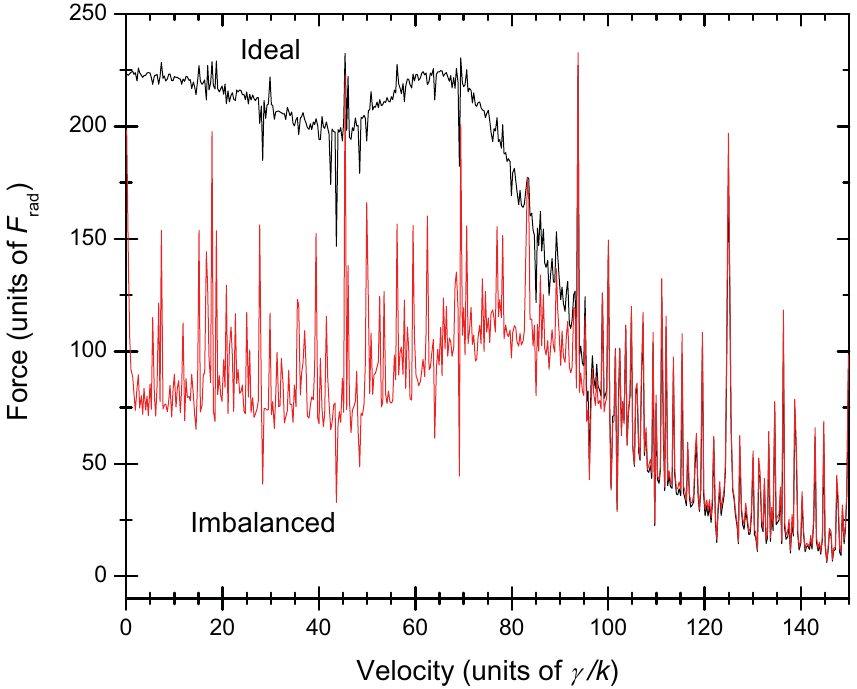}
\caption{(Color online) Four-color PCF velocity profiles for $\delta=250\gamma$ show a significant reduction of the force for a 25\% imbalance, but less than for the BCF in Fig. \ref{fig:BCF_ImbForceVsVelocity}, and this reduction largely disappears at velocities above 80 $\gamma/k$.}
\label{fig:PCF_ImbForceVsVelocity}
\end{figure}

\begin{figure}
\includegraphics[width=0.9\linewidth]{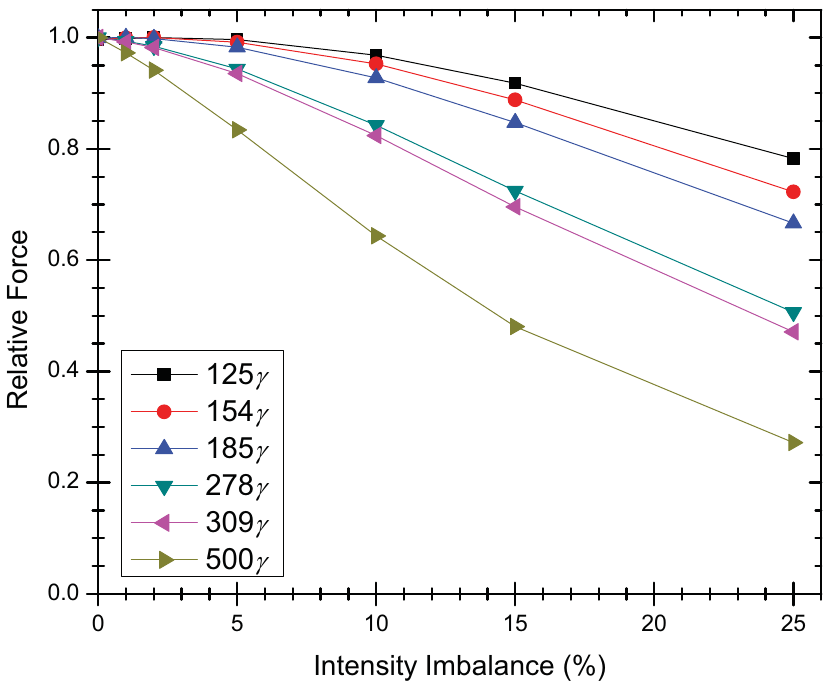}
\caption{(Color online) The four-color force is extremely robust against small beam intensity imbalances, although as the detuning increases, the force becomes increasingly sensitive to large imbalances.}
\label{fig:PCF_ForceVsImb}
\end{figure}

\begin{figure}
\includegraphics{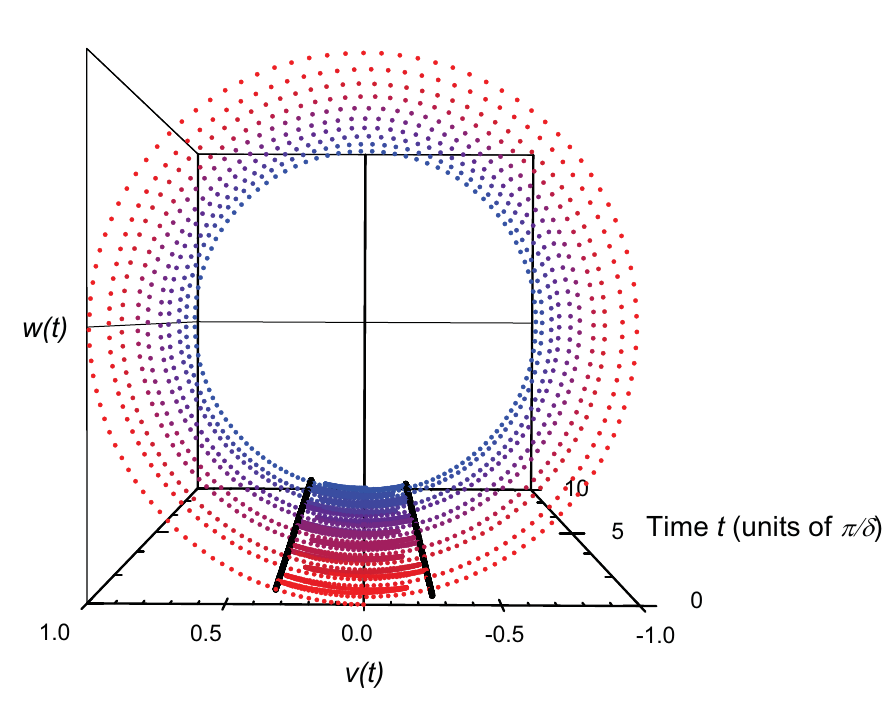}
\caption{(Color online) For the case of a 25\% beam imbalance, we see that the axis of symmetry again shifts away from the ground state. However, the symmetry shift is much less than for the BCF case shown in Fig. \ref{fig:BCF_ensembleImb}.}
\label{fig:PCF_ensembleImb}
\end{figure}

Polychromatic forces are also even more robust than the BCF against imbalances in left/right beam intensities. Figures \ref{fig:PCF_ImbForceVsVelocity}--\ref{fig:PCF_ensembleImb} show this behavior.  It is evident from Fig. \ref{fig:PCF_ForceVsImb} that there is almost no effect from imbalances up to 5\% except at extremely large detunings approaching 500~$\gamma$, which is outside the range used in experimental work to date.  At larger imbalances a gradual degradation is predicted.  Like the BCF case, the force reduction is probably underestimated for large imbalances at large detunings, because here it comes mostly from sharp multiphoton resonances, many of which are probably eliminated by decoherence under experimental conditions.
In the Bloch cylinder plot of Fig. \ref{fig:PCF_ensembleImb}, we note that a 25\% imbalance causes a noticeable shift of the axis of symmetry away from the ground state, but the shift is smaller than in Fig. \ref{fig:BCF_ensembleImb} and the total excursions are much smaller.  This likely accounts for the increased robustness of the four-color force.

We have also briefly investigated the impact of adding additional frequency components at $\pm 5 \delta$, and as one might expect, the benefits are not nearly so dramatic as for the initial additions at $\pm 3 \delta$.  Also, the addition of still more colors increases the likelihood that inadvertent transitions to distant states will compromise the desired two-level cycling system.  Finally, we have confirmed numerically that adding a second harmonic at $\pm 2 \delta$ is deleterious in all regards, because it destroys the symmetry of the odd-harmonic superposition.  Thus we see little reason to go beyond the four-color case except under unusual circumstances.

\subsection{Producing four-color beams \label{subsec:production}}

It might seem at first that a major obstacle in realizing the four-color PCF is the production of a laser beam with the required coherent superposition of four different frequencies.  Fortunately, this is easier than one might expect, because the frequencies are evenly spaced at intervals of $2\delta$.  They can be obtained by a slight variation of the standard scheme used to produce beams for BCF experiments, which utilizes a double-passed acousto-optic modulator (AOM) in which both the zero-order and first-order beams are retroreflected \cite{Williams00}. If the entering zero-order beam is at frequency $\omega-\delta$ and the acoustic frequency of the AOM is $2\delta$, the output consists of two beams that, if merged, contain equally-spaced frequencies at $\omega-3\delta$, $\omega-\delta$, $\omega+\delta$, and $\omega+3\delta$.  We plan to test this scheme with a metastable helium beam in our laboratory, where ample laser power will soon be available using tapered laser diode amplifiers that are rated to provide more than 1~W at 1083~nm.

  If may also be possible to generate the four-color beams by injecting rf radiation at multiples of $\delta$ into an electro-optic phase modulator (EOM) with amplitudes chosen to suppress the carrier.  However, an examination of the FM sideband spectrum for pure phase modulation indicates that at least three rf harmonics must be superposed to produce the correct pattern, and probably more, so the AOM-based scheme seems more appealing.

\subsection{Pulse Trains \label{subsec:pulseTrain}}

As stated previously, the basic concept behind the four-color force is that the addition of more colors creates a narrower pulse. In that spirit, why is it that we do not simply use a pulsed laser, or equivalently a chopped cw laser? This is indeed a possibility.  The first(and so far only) demonstration of stimulated forces on molecules used a train of mode-locked laser pulses~\cite{Voitsekhovich94}, and recently the group of Derevianko has proposed the use of a carefully tailored train of ultrashort $\pi$-pulses to produce stimulated cooling of molecules~\cite{Ilinova13}.

However, there are a few important limitations.  One is the purely technical issue that short-duration pulses with an area of $\sim$\,$\pi$ are not easily produced at arbitrary wavelengths.  A more basic concern is that short-duration pulses inherently have broad linewidths, extending across much of the visible spectrum for femtosecond pulses.  This will usually make it impossible to drive a pure two-level system, because numerous additional transitions will lie within the bandwidth.  The proposal in Ref. \cite{Ilinova13} is a special case that actually exploits the large bandwidth --- here, the frequency-domain ``teeth" of a phase-stabilized frequency comb are carefully matched to the rotational spacings of a molecule.

A more subtle but equally important consideration is that a simple pulse train lacks the robustness of polychromatic forces against left/right beam intensity imbalances. As explained in Sec. \ref{subsec:OBEsim}, it is the phase alternation between successive beat notes that gives rise to this robustness, because after a full four-pulse period (two pulses from the left, and two from the right), the net pulse area is zero even for imbalanced beams.  Pulse trains from a pulsed laser lack this phase alternation, and in fact they typically lack any phase coherence at all except in the case of a phase-stabilized frequency comb.  Thus any any imbalance between pairs of counterpropagating pulses will result in a residual phase on the Bloch sphere that accumulates with time, eventually reversing the direction of the force.  This problem has been confirmed in our numerical simulations by removing the sign alternation in the interference envelope. As expected, the sensitivity to imbalance then reverts to that of the simple $\pi$-pulse model as described in Section II.C of Ref. \cite{Chieda12}.

\section{Summary}

We have numerically studied the properties and behavior of systems under several different bichromatic and polychromatic light fields. With the assistance of ``Bloch cylinder" plots of the Bloch vector trajectory, we show that the surprising robustness of polychromatic forces with imbalanced beam intensities stems from sign alternation of the pulse areas in a train of beat notes.  This has important consequences for stimulated-force schemes relying on pulsed lasers.  We also analyze the excited-state fraction in detail, and show that for the ordinary BCF under optimal conditions, the time-averaged excited-state fraction is 41\%.

A proposed extension to four-color polychromatic forces shows great promise.  By adding components at $\pm3 \delta$ to the usual BCF beams at $\pm \delta$, the velocity range is increased by nearly a factor of three, the average excited-state fraction is reduced to 24\%, and the force is increased by nearly 50\%.  The total laser power is larger only by a factor of $4/3$, a modest price to pay compared with the factor of 9 that would be required for a BCF configuration at detuning 3$\delta$.  We also predict that the four-color force is significantly more robust in the presence of imbalanced beam intensities, especially if compared with a BCF with an increased detuning.

Plans are underway to test these ideas in metastable He, where a longitudinal declerator would benefit greatly from the increased velocity range. However, the greatest promise may be for applying stimulated optical forces to molecules, where a decreased excited-state fraction directly impacts the time available for the force to act prior to loss from radiative decay to dark states.

\begin{acknowledgments}
Funding for this research was provided by the NSF.
\end{acknowledgments}

\end{document}